\documentclass[draft,onecolumn]{IEEEtran}

\usepackage{color} 
\usepackage{amsfonts} 
\usepackage{amsmath} 
\usepackage{stmaryrd} 
\usepackage{amssymb} 
\usepackage{graphicx} 

\newcommand{\dl}{\llbracket}
\newcommand{\dr}{\rrbracket}
\newcommand*{\Scale}[2][4]{\scalebox{#1}{$#2$}}
\newcommand{\mat}[1]{\mbox{\boldmath{$#1$}}} 

\newtheorem{teorema}{Theorem}[section]
\newtheorem{lema}[teorema]{Lemma}

\newtheorem{prop}[teorema]{Proposition}
\newtheorem{cor}[teorema]{Corollary}

\begin{document}

\title{Generalized weights and bounds for error probability over erasure channels}


\author{Leandro~Cruvinel~Lemes and Marcelo~Firer,~\IEEEmembership{Member,~IEEE} \thanks{L.~C.~Lemes is with the Department of Electrical Engineering at Federal University of Triangulo Mineiro (UFTM) in Uberaba, MG, Brazil, e-mail: leandro@icte.uftm.edu.br}
\thanks{M. Firer is with Department of Mathematics at State University of Cam\-pi\-nas (UNICAMP), in Campinas, SP, Brazil, e-mail: mfirer@ime.unicamp.br}
\thanks{This paper was presented in part at 2014 Information Theory and Applications Workshop}}

\maketitle

\begin{abstract}
New upper and lower  bounds for the error probability over an erasure channel are provided, making use of Wei's generalized weights, hierarchy and spectra. In many situations the upper and lower bounds coincide and this allows improvement of existing bounds. Results concerning MDS and AMDS codes are deduced from those bounds.
\end{abstract}



\section{Introduction}

Generalized weights of a linear code were introduced by Victor Wei in \cite{wei} as a generalization of the minimal distance of a code. Wei's generalized weight  became relevant invariants in Coding Theory, being determined for
particular classes of codes (\cite{sti, feng, mun, bar, geo, hei, klove, 272498}) and bounded when explicit
formulas are not available (\cite{ash, dou1}). However, the importance of those
invariants concerning one of the main problems of Coding Theory - estimating
the efficiency of a code in terms of errors correction - has not yet been properly
explored. 

Considering a $q$-ary erasure channel we firstly give an expression for the error probability (Proposition \ref{eq8}) that separates the variables of the problem (namely, the code and the channel), where for error probability we mean either ambiguity probability or
the decoding error's probability. Considering the hierarchy and spectra of generalized weights, we are able to get new bounds for the error probability of linear codes (Theorem \ref{teoBOUNDS}). It turns out that, in many cases, the upper bounds for
ambiguity are better then the ones determined by Didier in \cite{did} and the lower bound better then those determined by Fashandi et al. (in \cite{Fas}, where the authors are concerned mainly with codes over large alphabets).  Recently, Liva, Paolini and Chiani (\cite{liva}) presented bounds for the error probability of a random code over $q$-ary erasure channels. The bounds are designed for codes with parity check matrix that are randomly generated and the results are shown to improve existing bounds for specific families of codes. In their approach, they use the weight distribution of the code to produce the bound for general codes. In some sense it is similar to the approach we adopt in this work, but we go further and consider not only the weight distribution but the spectra of the generalized weights.   

Having those bounds, we consider (Section \ref{MDS}) separation properties of a code (MDS and generalizations) and show that for MDS and AMDS codes the upper and lower bounds obtained for error probability collapse, becoming hence a closed expression for the error probability, (what was already known to Fashandi et. al \cite{Fas} in the MDS case). We conclude by showing the role of MDS and AMDS codes in minimizing the error probability when considering an erasure channel with overall error probability sufficiently small.

\section{Basic definitions and notation}

\subsection{Erasure Channel}
In this work we consider a \emph{Discrete Erasure Channel} (DEC) defined by an input alphabet $\mathcal{X}=\mathbb{F}_{q}$ (finite field with $q$ elements), an output alphabet ${\mathcal{Y}}={\mathbb{F}}_{q}\cup \{\epsilon \}$ (where 
$\epsilon \notin \mathbb{F}_{q}$ is called the \emph{erasure symbol}) and a probability function $\mathbb{P}_{i|j} := \text{Pr}\left[Y=j|X=i\right]$ defined by:
\begin{itemize}
\item[(a)] $\mathbb{P}_{j|i} = 0$, for $i\neq j$ and $\{i,j\} \subseteq {\cal{X}}$;
\item[(b)] $\mathbb{P}_{i|i} = 1-\mathtt{p}$, for $0\leq \mathtt{p} \leq 1$;
\item[(c)] $\mathbb{P}_{\epsilon|i} = \mathtt{p}$, for $i\in {\cal{X}} ,$
\end{itemize}
The constant $0<\mathtt{p}<1/2$ is called the \emph{overall error probability} of the channel.

A \emph{Discrete Memoryless Erasure Channels} (DMEC) is obtained by defining 
\begin{equation}\label{mvs}
P(\mathbf{y}|\mathbf{x})=\prod_{l=1}^{n}\mathbb{P}_{y_l|x_l},
\end{equation}
where $P(\mathbf{y}|\mathbf{x})$ is the probability that a message $\mathbf{y}$ is received given that $\mathbf{x}$ was sent,  $\Scale[1]{\mathbf{x}=(x_1,x_2,\cdots ,x_n )\in {\cal{X}}^n} $ and $\Scale[1]{\mathbf{y}=(y_1,y_2,\cdots ,y_n)\in {\cal{Y}}^n}$.


\subsection{Generalized weights}

Given integers $r,s\in\mathbb{Z}$ with $r<s$, we denote $\llbracket r,s\rrbracket :=\{r,r+1,\ldots,s-1,s\}$. For simplicity, we write $\llbracket n\rrbracket:=\llbracket 1,n\rrbracket$. 

From here on we assume that   $C\subseteq \mathbb{F}^{n}_{q}$ is an $[n,k]_q$-linear code. 
Given $\mathbf{x}\in {\mathbb{F}}^{n}_{q}$, the \textit{support} of $\mathbf{x}$ is $$\text{supp}(\mathbf{x}) = \{i \in \dl n \dr; \text{with}\ x_i\neq 0\}\text{.}$$
The Hamming weight and distance may be expressed counting the supports: $w(\mathbf{x})=\left| \text{supp}(\mathbf{x})\right|$ and $d(\mathbf{x},\mathbf{\tilde{x}})= \left| \text{supp}(\mathbf{x}-\mathbf{\tilde{x}})\right|$ respectively, where $\left| A \right|$ denotes the cardinality of $A$.
Given a subcode $D\subseteq C$, the \textit{support of} $D$ is defined as 
$$\text{supp}(D) = \bigcup_{\mathbf{x}\in D}\text{supp}(\mathbf{x})$$
and the \textit{generalized weight} $d_i(C)$ of $D$ is defined as
$$d_{i}(C) = \min \{|\text{supp}(D)|; D\subseteq C\ \text{and}\ \dim\left(D\right)=i\},$$
for $i\in \dl k\dr$. 
Those weights generalize the Hamming weight, in the sense that $d_1(C)$ is the usual minimal distance $d(C)$.
 
It is well known (Wei's Monotonicity Theorem \cite{wei}) that the generalized weights are strictly increasing 
\[ \label{teoMONO}
d_{1}(C) <d_{2}(C) < \cdots < d_{k}(C)
\]
and we call $\left\{ d_{1}(C),d_{2}(C),\cdots ,d_{k}(C)\right\}$ the \textit{weight hierarchy of} $C$.

We denote by  ${\cal{A}}^{i}_{j} = {\cal{A}}^{i}_{j}(C)$, $i\in \dl 0, k\dr$ and $j \in \dl 0, n\dr$, the set of all $i$-dimensional linear subcodes $D \subseteq C$ supported by $j$ coordinates, that is,
$${\cal{A}}^{i}_{j}(C) = \{D \subseteq C; |\text{supp}(D)|=j\ \text{and}\ \dim\left(D\right)=i\}.$$
The cardinality of ${\cal{A}}^{i}_{j}$ is called the $i$-th \textit{generalized spectra with support } $j$ of the code $C$ and we denote 
$$A_{j}^{i} = |{\cal{A}}^{i}_{j}|.$$ We call the matrix $\left(A^i_j\right)_{i=0,\cdots ,k; \\ j=0,\cdots n}$ the \textit{spectra-matrix of the code}.


\subsection{Ambiguity}

Considering a DMEC and given a code $C\subseteq \mathbb{F}_q^n$, some messages in ${\cal{Y}}^n$ can never be received. We denote by $E_{C}$ the subset of messages that may be received, that is,  
\begin{equation*}
E_{C}=\{\mathbf{y}\in {\mathcal{Y}}^{n};\mathbb{P}_{\text{receive}}(\mathbf{y})\neq 0\}
\end{equation*}
and call it the set of \textit{admissible} or \textit{possible messages}, where $\mathbb{P}_{\text{receive}}(\mathbf{y})$ is the probability to receive $\mathbf{y}\in {\cal{Y}}^n$.

Given $\mathbf{x}\in {\cal{X}}^{n}$, we denote by $\mathbb{P}_{\text{send}}(\mathbf{x})$ the priori probability of $\mathbf{x}$.

Since ${\mathbb{P}}_{\text{receive}}(\mathbf{y}) = \sum_{\mathbf{x}\in C}P(\mathbf{y}|\mathbf{x})\mathbb{P}_{\text{send}}(\mathbf{x})$, from expression  (\ref{mvs}) it follows that
 
\begin{equation*}
E_{C}=\{\mathbf{y}\in {\mathcal{Y}}^{n};\exists \mathbf{x}\in C\ \text{with}\ \mathbb{P}_{\text{send}}(\mathbf{x})\prod_{l=1}^{n}\mathbb{P}_{y_l|x_l} \neq 0\}.
\end{equation*}


Given $\mathbf{y}\in {\cal{Y}}^{n}$ let $R:=R(\mathbf{y})=\{i\in \dl n\dr; y_i = \epsilon \}$ and define the  \textit{set} $[\mathbf{y}]_R$ \textit{of} $R$\textit{-ambiguities} of $\mathbf{y}$ as
$$[\mathbf{y}]_R:= \{\mathbf{x}\in C; P(\mathbf{y}|\mathbf{x})\neq 0\}.$$ Despites the notation, the set $[\mathbf{y}]_R$ depends both on $\mathbf{y}$ and $C$.  Using a Maximum Likelihood decoder, once the message $\mathbf{y}$ is received, the elements of $[\mathbf{y}]_R$ are the possible choices for decoding $\mathbf{y}$. A vector $\mathbf{y}$ is said to be an \textit{ambiguity} of $C$ if $|[\mathbf{y}]_R| > 1$.

Identifying $\mathbb{F}_q^n$ with the product  $\Pi_{i=1}^{n}\left(\mathbb{F}_q\right)_i$, given $R\subseteq \dl n \dr$,  $\pi_R$ denotes the projection of $\mathbf{x} = \left(x_i\right)_{i\in \dl n \dr}$ in the coordinates of $R$: $\pi_R(\mathbf{x}) = \left(x_i\right)_{i\in R}$. To shorten the notation we will write $\pi_R(\mathbf{x}) = {\mathbf{x}}^{R}$. Denote by  $E_R$ the 
set of admissible messages with erasures on the coordinates in $R$, that is, 

$$E_R = \{\mathbf{y}\in E_C; y_i=\epsilon,\ \forall i\in R\}.$$

Given $R\subseteq \dl n \dr$ let $\bar{R}:=\left\{i\in \dl n \dr; i\notin R\right\}$ be the \textit{complement of} $R$ (\textit{in} $\dl n \dr$). 

The following proposition consists of a sequence of elementary properties that are stated for future reference.

\begin{prop}\label{proposicao1} Considering a DMEC,  let $C$ be a linear code, $\mathbf{y} \in {\mathcal{Y}}^{n}$ and $R = R(\mathbf{y}) = \{i\in \dl n \dr; y_i = \epsilon\}$. Then:
\begin{enumerate}
\item[(i)] \label{lema1}
$\mathbf{y}\in E_C$ \emph{iff} $[\mathbf{y}]_R  \neq \emptyset$  ; \vspace{0.2cm} 
\item[(ii)] \label{lema2}
$[\mathbf{y}]_R = \{\mathbf{x}\in C; {\mathbf{x}}^{\bar{R}}=\mathbf{y}^{\bar{R}}\}$; \vspace{0.2cm}
\item[(iii)] \label{lema3}
$[\mathbf{0}]_R$ is the kernel of the projection map $\pi_{\bar{R}}$ restricted to the code $C$; \vspace{0.2cm}
\item[(iv)] \label{lema4}
For any $\mathbf{y}\in E_C$,  $|[\mathbf{y}]_R|= |[\mathbf{0}_R]|$; \vspace{0.2cm}
\item[(v)] \label{lema5}
$|E_R|=q^k|[\mathbf{0}]_R|^{-1}$.
\end{enumerate}
\end{prop}

\begin{IEEEproof}

Statements  (i), (ii) and (iii) follow trivially from the definitions. 

To prove item (iv), consider an admissible message $\textbf{y} \in E_{C}$ with 
$$R = R(\mathbf{y}) = \{i\in \dl n \dr; y_i = \epsilon\}.$$  Item (i) ensures that $[\textbf{y}]_{R}\neq \emptyset$, so it is possible to choose (and fix) an element $\mathbf{c_0} \in [\textbf{y}]_{R}$. Fixed $\mathbf{c_0}$, define the map $\phi := \phi_{\mathbf{c_0}}$ 
$$\begin{matrix}
\phi: & [\textbf{y}]_{R} & \longrightarrow & [\textbf{0}]_{R}\\
& \textbf{c} & \longmapsto & \textbf{c} + (q-1)\mathbf{c_0}.
\end{matrix}$$
First of all, we remark that 
$$(\textbf{c} + (q-1)\mathbf{c_0})^{\bar{R}}= \textbf{c}^{\bar{R}} + (q-1)\mathbf{c_0}^{\bar{R}}= \mathbf{c_0}^{\bar{R}} + (q-1)\mathbf{c_0}^{\bar{R}}=\textbf{0}^{\bar{R}},$$
for any $\textbf{c}\in [\textbf{y}]_R$. So that $\phi$ is well defined. To establish that $|[\mathbf{y}]_R|= |[\mathbf{0}_R]|$ one should prove that $\phi$ is a bijection. If $\phi(\mathbf{c_1})=\phi(\mathbf{c_2})$ then $\mathbf{c_1} + (q-1)\mathbf{c_0} = \mathbf{c_2} + (q-1)\mathbf{c_0}$, hence $\mathbf{c_1} = \mathbf{c_2}$, so that  $\phi$ is injective. 
To prove that $\phi$ is surjective, let $\textbf{c} \in [\textbf{0}]_{R}$. 
Since $\mathbb{F}_q=\mathcal{X}$ is a vector field with $q$ elements, it follows that 
\[
(\textbf{c} + \mathbf{c_0}) + (q-1)\mathbf{c_0}= \textbf{c} + q\mathbf{c_0} = \textbf{c}
\]
so, if we can show that $\textbf{c} + \mathbf{c_0}\in [\mathbf{y}]_{R}$ it will follow that $\phi(\textbf{c} +\mathbf{c_0})=\textbf{c}$.

Since  $\textbf{c} \in [\textbf{y}]_{R}$ it follows that $\textbf{c}^{\bar{R}}=\textbf{0}^{\bar{R}}$and since 
\begin{eqnarray*}
(\mathbf{c_0}+\textbf{c})^{\bar{R}}=\mathbf{c_0}^{\bar{R}}+\textbf{0}^{\bar{R}}=\mathbf{c_0}^{\bar{R}}
\end{eqnarray*}
we get that $\mathbf{c_0}+\textbf{c} \in [\textbf{y}]_{R}$, hence $\phi$ is a bijection and $|[\mathbf{y}]_R|= |[\mathbf{0}_R]|$.

To prove item (v), consider $R \subseteq \llbracket n\rrbracket$ .The code $C$ may be expressed as the union 
$$C = \bigcup_{\textbf{y}\in E_{R}}[\textbf{y}]_{R},$$
and since for $\textbf{x}\neq \textbf{y}$, $\textbf{x}, \textbf{y} \in E_R$, it follows that  $$[\textbf{x}]_{R}\cap [\textbf{y}]_{R} = \emptyset,$$ and hence this union is disjoint.

It follows that 
$$q^{k}= |C|= \sum_{\textbf{y}\in E_{R}}|[\textbf{y}]_{R}|.$$
But item (iv) ensures that $|[\mathbf{y}]_R|= |[\mathbf{0}_R]|$ so

$$\sum_{\textbf{y}\in E_{R}}|[\textbf{y}]_{R}|=\sum_{\textbf{y}\in E_{R}}|[\textbf{0}]_{R}|=|E_{R}||[\textbf{0}]_{R}|,$$
and the statement in item (v) is true.
\end{IEEEproof}

\subsection{Error probability for ambiguity and decoding}

We are considering an DMEC with conditional probabilities defined by (\ref{mvs}), with overall error probability $\mathtt{p}$. Given a code $C$ we assume that the prior probability is identically distributed on $C$, that is, ${\mathbb{P}}_{\text{send}}(\mathbf{x})=\left| C\right|^{-1}$, for any $\mathbf{x}\in C$.

Given $\mathbf{y}\in {\cal{Y}}^n$,  denote 
by $\mathbb{P}_{\text{amb}}(\mathbf{y})$ the probability that $\mathbf{y}$ is ambiguous. The \textit{ambiguity probability of } an $[n,k]_q$-code $C$ (the error probability before any decoding procedures is produced) is 
\begin{align*}
P_{\text{amb}}(C) &= \sum_{\mathbf{y}\in {\cal{Y}}^n}\mathbb{P}_{\text{amb}}(\mathbf{y})\mathbb{P}_{\text{receive}}(\mathbf{y}) \\
&= \sum_{\mathbf{y}\in E_C}\mathbb{P}_{\text{amb}}(\mathbf{y})\mathbb{P}_{\text{receive}}(\mathbf{y}),
\end{align*} 
where the last equality follows from statement (i) in Proposition \ref{proposicao1}.

Considering a  maximum likelihood decoding criteria, denote by $\mathbb{P}_{\text{dec}}(\mathbf{y})$ the probability of $\mathbf{y}\in {\cal{Y}}^n$ being decoded incorrectly and define the \textit{decoding error probability of} an  $[n,k]_q$-code $C$ as
\begin{align*}
P_{\text{dec}}(C) &= \sum_{\mathbf{y}\in {\cal{Y}}^n}\mathbb{P}_{\text{dec}}(\mathbf{y})\mathbb{P}_{\text{receive}}(\mathbf{y})\\
&= \sum_{\mathbf{y}\in E_C}\mathbb{P}_{\text{dec}}(\mathbf{y})\mathbb{P}_{\text{receive}}(\mathbf{y}),
\end{align*}
where the last equality again follows from statement (i) in Proposition \ref{proposicao1}.

We  use  $P_{\ast}(C)$ to denote either $P_{\text{amb}}(C)$ or $P_{\text{dec}}(C)$, that is, we may consider $\ast$ to mean either `$\text{dec}$' or  `$\text{amb}$', so that both the previous expressions may be written as 

\begin{equation}\label{PdecPamb}
P_{\ast}(C) = \sum_{\mathbf{y}\in E_C}\mathbb{P}_{\ast}(\mathbf{y})\mathbb{P}_{\text{receive}}(\mathbf{y}).
\end{equation}

Assuming that ${\mathbb{P}}_{\text{send}}(\mathbf{x})=q^{-k}$, for any $\mathbf{x}\in C$, the probability that a message $\mathbf{y}$ is received is 
\begin{align*} 
\mathbb{P}_{\text{receive}}(\mathbf{y})&= \sum_{\mathbf{x}\in C}P(\mathbf{y}|\mathbf{x})\mathbb{P}_{\text{send}}(\mathbf{x}) \\
&= \frac{1}{q^k}\sum_{\mathbf{x}\in C}P(\mathbf{y}|\mathbf{x}).
\end{align*}

Considering an  admissible message $\mathbf{y}\in E_C$ and $R = R(\mathbf{y}) = \{i\in \dl n \dr; y_i = \epsilon\}$, it follows that
\begin{align*}
{\mathbb{P}}_{\text{receive}}(\mathbf{y})&= \sum_{\mathbf{x}\in [\mathbf{y}]_R}q^{-k}{\mathtt{p}}^{|R|}\left(1-\mathtt{p}\right)^{|\bar{R}|}\\ &=\sum_{\mathbf{x}\in [\mathbf{0}]_R}q^{-k}{\mathtt{p}}^{|R|}\left(1-\mathtt{p}\right)^{|\bar{R}|}\\ &=|[\mathbf{0}]_R|q^{-k}{\mathtt{p}}^{|R|}\left(1-\mathtt{p}\right)^{|\bar{R}|}.
\end{align*}


Substituting into equation (\ref{PdecPamb}) we get
\begin{equation}\label{acima1}
P_{\ast}(C) = \sum_{\mathbf{y}\in E_C}\mathbb{P}_{\ast}(\mathbf{y})|[\mathbf{0}]_R|q^{-k}{\mathtt{p}}^{|R|}\left(1-\mathtt{p}\right)^{|\bar{R}|},
\end{equation}
hence  (\ref{acima1}) may be expressed as
\begin{equation*}
P_{\ast}(C) = \sum_{R\subseteq \dl n \dr}\mathbb{P}_{\ast}(\mathbf{y})|E_R||[\mathbf{0}]_R|q^{-k}{\mathtt{p}}^{|R|}\left(1-\mathtt{p}\right)^{|\bar{R}|}
\end{equation*}
and, from statement (v) in Proposition \ref{proposicao1} it follows that
\begin{equation}\label{expPast}
P_{\ast}(C) = \sum_{R\subseteq \dl n \dr}\mathbb{P}_{\ast}(R){\mathtt{p}}^{|R|}\left(1-\mathtt{p}\right)^{|\bar{R}|}.
\end{equation}
Note that comparing expressions (\ref{acima1}) and (\ref{expPast}), $\mathbb{P}_{\ast}(\mathbf{y})$ was replaced by $\mathbb{P}_{\ast}(R)$ and this is possible since those probabilities do not depend on ${\mathbf{y}}$ but only at what are the erased coordinates of ${\mathbf{y}}$, that is, on the set $R$.

Denoting
\begin{equation}
Q_{\ast,r}= \sum_{R; |R|=r}P_{\ast}(R),
\end{equation}
we may write (\ref{expPast}) as
\begin{equation}\label{expancao_Q}
P_{\ast}(C) =\sum_{r=0}^{n} Q_{\ast,r} \mathtt{p}^{r} (1-\mathtt{p})^{n-r}.
\end{equation}
We note that 
\begin{equation}\label{defQdec}
Q_{dec,r} =\sum_{\{R\subseteq \dl n\dr;|R|=r\}}{\left( 1-\frac{1}{|[\mathbf{0}]_{R}|}\right)}. 
\end{equation}
\begin{equation}\label{defQamb}
Q_{amb,r} =\sum_{\{R\subseteq \dl n\dr;|R|=r\}}{\left\lceil 1-\frac{1}{|[\mathbf{0}]_{R}|}\right\rceil }.
\end{equation}
where  $\left\lceil \alpha \right\rceil$ is the smallest integer greater or equal to $\alpha \in {\mathbb{R}}$, hence  
\[ \left\lceil 1-\frac{1}{|[\mathbf{0}]_{R}|}\right\rceil = \left\{ \begin{array}{ll}
         0  & \mbox{if $|[\mathbf{0}]_{R}|=1$};\\
        1 & \mbox{if $|[\mathbf{0}]_{R}|>1$},\end{array} \right. \]
that is, it equals $1$ or $0$ according if there is more then one or only one (namely $\mathbf{y}$) admissible messages having $R$ as a set of ambiguous coordinates.

We define 
\begin{equation}\label{airnew}
a^i_r:=a^i_r(C)=|\{R\in \dl n\dr; |R|=r\ \text{and}\ \dim([\mathbf{0}]_{R})=i\}|
\end{equation}
and 
\[ \delta_{\ast,i} = \left\{ \begin{array}{ll}
         1- \frac{1}{q^{i}}, & \mbox{for $\ast=\text{dec}$};\\
        \left\lceil 1- \frac{1}{q^{i}}\right\rceil, & \mbox{for $\ast = \text{amb}$}.\end{array} \right. \]
Since $[\mathbf{0}]_{R}$ is the kernel of the projection $\pi_{\bar{R}}$ restricted to $C$, it follows that each $a^i_r$ depends on the code $C$, . We also remark that \begin{equation}\label{sumsum}
\sum^{k}_{i=0}a^i_r = \binom{n}{r}.
\end{equation}

Using this notation it is possible to write 
\begin{equation}\label{defqast}
Q_{\ast,r} = \sum_{i=0}^{k}a^i_r\delta_{\ast,i}
\end{equation}
so that equation (\ref{expPast}) may be expressed in a vectorial form, as follows:
\begin{prop} \label{eq8} The ambiguity probability and the decoding error probability of a linear code may be expressed as the product
\begin{equation}\label{principal}
P_{\ast}(C) = \mat{\delta}_{\ast}\Lambda {\mat{\rho}}^{T}
\end{equation}
where 
$$\mat{\delta}_{\ast} = \left(\delta_{\ast,0},\ldots, \delta_{\ast,k}\right),$$
$$\mat{\rho} = \left( (1-\mathtt{p})^{n} , \mathtt{p}(1-\mathtt{p})^{n-1}, \ldots, \mathtt{p}^{n}\right),$$ 
$\mat{\rho}^T$ is the transpose of the vector $\mat{\rho}$ and
\begin{equation}\label{matrizlambda} 
\Lambda = \left( \begin{array}{llll}
      a_{0}^{0} & a_{1}^{0} & \cdots & a_{n}^{0}\\
			\vdots & \vdots &  & \vdots \\
			a_{0}^{k} & a_{1}^{k} & \cdots & a_{n}^{k} \end{array} \right).
\end{equation}
\end{prop}
It is important to remark that  whether $\ast$ means ``decoding"  or ``ambiguity", $\mat{\delta}_{\ast}$ depends only on the parameters $\left[n,k\right]_q$; $\mat{\rho}$  depends only at the channel not on the code; the matrix  $\Lambda$ depends only on the code $C$, not on the channel neither on $\ast$ meaning ``decoding"  or ``ambiguity". The matrix $\Lambda$ is called the \textit{support-matrix of} $C$ and since it is the only factor in equation (\ref{principal}) that depends on the code $C$, bounds for $P_{\ast}(C)$  will be produced by focusing  the attention on this matrix.

\section{Support-matrix and spectra-matrix of a code}

The goal of this section is to establish a relation between the support-matrix and the spectra-matrix of a code, a relation which will be the key to establish new bounds for $P_{\ast}$. 

From here on,  without loss of generality, it is assumed that $d_k = n$. We start with two simple lemmas.

\begin{lema}\label{lemax}
Let $C$ be a code, $D\in {\cal{A}}^{i}_{r}(C)$ and $s\in \dl n\dr$. Then, $D$ contains a set of linearly independent vectors $\{\mathbf{x}^{1},\ldots, \mathbf{x}^{i-1}\}\subset D$ such that $\pi_{s}(\mathbf{x}^{j})=0$, for any $j\in \dl i-1 \dr$.
\end{lema}

\begin{IEEEproof}
Let $R=supp(D)$ and suppose that $s\in R$. Since $R=supp(D)$, there is $\mathbf{x} \in D$ such that $\pi_{\{s\}}(\mathbf{x})\neq 0$. Consider any ordered basis  of the subspace $D$ containing $\mathbf{x}$, let us say $\{\mathbf{x},
\mathbf{u}^{1}\ldots,\mathbf{u}^{i-1}\}$. Defining
\[
 \mathbf{x}^{j}= \mathbf{u}^{j}-\frac{\pi_s(\mathbf{u}^{j})}{\pi_s(\mathbf{x})}\mathbf{x}\text{,}
\]
it is immediate to check that $\{\mathbf{x}^{1},\ldots, \mathbf{x}^{i-1}\}$ is linearly independent and $\pi_{s}(\mathbf{x}^{j})=0$, for any $j\in \dl i-1 \dr$.
To conclude the proof, let us assume that  $s\notin R$. This implies that $\pi_{\{s\}}(\mathbf{x})=0$, for every $\mathbf{x}\in D$ and the result follows from the fact that $\dim(D)=i$.
\end{IEEEproof}

\begin{lema}\label{L32}
For every $i\in \dl k \dr$ the coefficient $A^{i}_{d_i}$ of the spectra-matrix depends on the number of different supports  attained by subcodes in ${\cal{A}}^{i}_{d_i}$, that is, 
$$A^{i}_{d_i}=|\{R; R=\text{supp}\left(D\right)\ \text{and}\ D\in {\cal{A}}^{i}_{d_i}\}|.$$
\end{lema}

\begin{IEEEproof}
It is enough to prove that given $D_1,D_2\in {\cal{A}}^{i}_{d_i}$, if $D_1\neq D_2$, then  $supp\left(D_1\right)\neq supp\left(D_2\right)$.

Since $D_1\neq D_2$ it is possible to assume wlog there is $\mathbf{x}\in D_{2}\setminus D_{1}$. Suppose  $supp(D_{1})=supp(D_{2})=R$ and we will show that this leads to a contradiction. Since $\mathbf{x}\in D_{2}\setminus D_{1}$, the subspace $D=\left\langle \{\mathbf{x}\} \cup D_{1}\right\rangle$ has dimension $i+1$ and since $supp(\mathbf{x})\subseteq supp(D_{1})=R$, it follows that  $supp(D)=R$. From Lemma \ref{lemax}, given $s\in R$ there is a linearly independent set $\{ \mathbf{x}^{1},\ldots, \mathbf{x}^{i} \}\subseteq D$ such that 
$\pi_{\{s\}}(\mathbf{x}^{j})=0$, $\forall j\in \dl i\dr$. Considering the subspace $\tilde{D}=\left\langle \mathbf{x}^{1},\ldots, \mathbf{x}^{i}\right\rangle$ it is an $i$-dimensional subspace of $C$ with $supp(\tilde{D}) \subseteq R \setminus \{s\}$, that is, $|supp(\tilde{D})| < d_{i}$, contradicting the minimality of $d_i$.
\end{IEEEproof}

Now it is possible to determine $a^i_r = 0$ for $r\leq d_{i}$:
\begin{prop}\label{P33} For any $i=1,\cdots k$, the coefficients of the support-matrix of a code $C$ satisfy 
\[ a^i_r = \left\{ \begin{array}{ll}
          0  &\mbox{ for $r<d_i$};\\
         A^i_{d_i}   &\mbox{ for $r=d_i$}.\end{array} \right. \]
\end{prop}

%

\begin{IEEEproof}
We first consider the case $r=d_i$. Defining
$$B := \{R\subseteq \dl n\dr; \dim\left([\mathbf{0}]_{R}\right)=i\ \text{and}\ |R|=d_{i}\}\text{.}$$
we have, by definition, that $|B|=a^{i}_{d_{i}}$. The Lemma \ref{L32} ensures that it is sufficient to prove that 
$$B = \{R; R=\text{supp}\left(D\right)\ \text{and}\ D\in {\cal{A}}^{i}_{d_i}\}.$$
Let us consider $R\subseteq \llbracket n\rrbracket$ with  $\dim([\mathbf{0}]_{R})=i$ and $|R|=d_{i}$ and let us prove that $supp([\mathbf{0}]_{R})=R$. The definition of $[\mathbf{0}]_{R}$ ensures that $supp([\mathbf{0}]_{R})\subseteq R$ and so $|supp([\mathbf{0}]_{R})|\leq |R|$. Since $\dim([\mathbf{0}]_{R})=i$ and $|supp([\mathbf{0}]_{R})|\leq |R|= d_{i}$, it follows that  $supp([\mathbf{0}]_{R})= R$ hence $a^{i}_{d_{i}}=A^{i}_{d_{i}}$.

We consider now the case $r<d_i$. Given $r<d_i$, suppose that $a^{i}_{r}>0$. This implies there is $R\subseteq \dl n\dr$ with $|R|=r$ and $\dim([\mathbf{0}]_{R})=i$. But $supp([\mathbf{0}]_{R})\subseteq R$ and this implies $r = |R|\geq |supp([\mathbf{0}]_{R})|\geq d_{i}$, a contradiction. It follows that  $a^i_r=0$, for all $r<d_{i}$.
\end{IEEEproof}

We remark that the condition $r=d_i$ is strictly necessary in Proposition \ref{P33}. Considering for example the $[3,2]_2$-code generated by the vectors $\{(1,0,0),(0,1,1)\}$ for $r=2$, $d_1=1$ and $d_2=3$ we have that $A^1_2 = 1$ and $a^{1}_{2}=3$.
 
Wei's Monotonicity Theorem states that $i<j$ implies $d_i<d_j$, so Proposition \ref{P33} ensures the following:

\begin{cor}\label{pequeno}
If $j<i$ then  $a^{i}_{d_j}=0$.
\end{cor}

We continue with some results that will be used to produce the expected bounds for the error probability.

\begin{lema}\label{L33}
Let $R \subsetneq \dl n\dr$, with $|R|=r$ and $i=\dim([\mathbf{0}]_{R})$. Then, 
\begin{equation}  \label{eql2t3}
\dim([\mathbf{0}]_{R\cup \{j\}})\geq \max \{k+r+1-n,i\}\text{,}
\end{equation}
for any $j\in \dl n\dr\setminus R$.
\end{lema}

\begin{IEEEproof}
Denoting $S_j = R\cup\{j\}$, from item (iii) in Proposition \ref{proposicao1} it follows that $[\mathbf{0}]_{S_j}= \ker(\pi_{\bar{S_j}})$. Considering that $\pi_{\bar{S_j}}$ may be expressed as the composition 
\begin{equation}\label{p3q1}
\pi_{\bar{S_j}}= \pi_{\overline{\{j\}}}\circ \pi_{\bar{R}}
\end{equation}
of the projections 
$$\begin{matrix}
C & \stackrel{\pi_{\bar{R}}}{\longrightarrow} & {\mathbb{F}}^{n-r}_{q} & \stackrel{\pi_{\overline{\{ j\}}}}{\longrightarrow} & {\mathbb{F}}^{n-r-1}_{q}\\
\sum_{s\in \dl n\dr}c_{s}\mathbf{e}_{s} & \longmapsto & \sum_{s\in \bar{R}}c_{s}\mathbf{e}_{s} & \longmapsto & \sum_{s\in \bar{S_j}}c_{s}\mathbf{e}_{s},
\end{matrix}$$
the classical Kernel Theorem ensures that 
$$
\dim(C)=\dim(Im(\pi_{\bar{S_j}}))+\dim(\ker(\pi_{\bar{S_j}})),
$$
that is,
\begin{equation}\label{p3q2}
\dim\left([\mathbf{0}]_{S_j}\right)=k - \dim(Im(\pi_{\bar{S_j}})).
\end{equation}
But equation (\ref{p3q1}) implies
\begin{equation}\label{p3q3}
\dim(Im(\pi_{\bar{S_j}})) \leq \min\{\dim(Im(\pi_{\bar{R}})),\dim(Im(\pi_{\bar{\{j\}}}))\}.
\end{equation}
Since $\pi_{\bar{\{j\}}}$ determines a projection of an $(n-r)$-dimensional space into an $(n-r-1)$-dimensional subspace, it follows that 
\begin{equation}\label{p3q4}
\dim(Im(\pi_{\overline{\{j\}}}))=n-r-1
\end{equation}
and since $\dim([\mathbf{0}]_{R})=\dim(\ker(\pi_{\bar{R}}))$, we have that
\begin{equation}\label{p3q5}
\dim(Im(\pi_{\bar{R}})) = \dim(C)- \dim(\ker(\pi_{\bar{R}}))=k-i.
\end{equation}
It follows from (\ref{p3q3}), (\ref{p3q4}) and (\ref{p3q5}) that 
\begin{equation}\label{p3q6}
\dim(Im(\pi_{\bar{S_j}})) \leq \min\{k-i,n-r-1\}
\end{equation}
and equations (\ref{p3q2}) and (\ref{p3q6}) together imply
\begin{align*}
\dim([\mathbf{0}]_{S_j}) & \geq k - \min\{k-i,n-r-1\} \\ &= \max\{i,k+r+1-n\}.
\end{align*}
\end{IEEEproof}

The next propositions will be used to establish the bounds in Theorem \ref{teoBOUNDS} and both follow  from   Lemma \ref{L33}.

\begin{cor}\label{c8t2}
If $C$ is  an $[n,k]_{q}$-linear code $C$ then  $|[\mathbf{0}]_{R}|\geq q^{k-n+r}$ for every subset $R\subseteq \llbracket n\rrbracket$ with $|R|=r$, .
\end{cor}

\begin{IEEEproof}
The proof is made by induction on $|R|$. For the initial step, $|R|=0$, the result is satisfied since $R=\emptyset$. Suppose $|[\mathbf{0}]_{R}|\geq q^{k-n+|R|}$ for every $R\subseteq \llbracket n\rrbracket$ with $|R|\leq r$ and let us prove it also holds for $J\subseteq \llbracket n\rrbracket$ with $|J|=r+1$. We write  $J=R\cup\{j\}$ with $|R|=r$ and $j\notin R$, and from Lemma \ref{L33} it follows that 
\begin{eqnarray*}
\dim([\mathbf{0}]_{J})=\dim([\mathbf{0}]_{R\cup\{j\}})\geq \max \{k+r+1-n,i\}
\end{eqnarray*}
with $i=\dim([\mathbf{0}]_{R})$.
The induction hypothesis implies  that $i\geq k-n+r$ hence
\begin{align*}
\dim([\mathbf{0}]_{J})&\geq \max \{k+r+1-n,k-n+r\}\\ &=k-n+(r+1).
\end{align*}
\end{IEEEproof}

\begin{prop}\label{P35} If $r\geq n-k+i+1$, then $a^i_r=0$.
\end{prop}

\begin{IEEEproof}
Suppose $a^i_r>0$ for some $r\geq n-k+i+1$, that is, suppose there  is a set $R\subseteq \llbracket n\rrbracket$ with  $\dim([\mathbf{0}]_{R})=i$ and 
\begin{equation}\label{eq456}
|R|:=r\geq n-k+i+1.
\end{equation}
We cannot have $R= \llbracket n\rrbracket$, since this would imply $r=n$, $[\mathbf{0}]_R=C$ and $i=k$, contradicting inequality (\ref{eq456}). So, let us assume that $R\subsetneq \dl n\dr$, so there is $j\in \dl n\dr\setminus R$. From Lemma \ref{L33} it follows that
$$\dim([\mathbf{0}]_{R\cup \{j\}})\geq \max \{k+r+1-n,i\}$$
and  inequality (\ref{eq456}) implies $k+r+1-n \geq i+2 >i$, hence
$$\dim([\mathbf{0}]_{R\cup \{j\}})\geq i+2\text{.}$$
It follows there is a subcode  $D\subseteq [\mathbf{0}]_{R\cup\{j\}}$ such that  $\dim (D)=i+2$ and Lemma \ref{lemax} ensures the existence of a subcode $\tilde{D} \subseteq D$ such that $\dim(\tilde{D})=i+1$ and $supp(\tilde{D})\subseteq R$. But $\tilde{D}\subseteq [\mathbf{0}]_{R}$ and $\dim([\mathbf{0}]_{R})=i$, a contradiction and so, for $r\geq n-k+i+1$, there is no $R\subseteq \dl n\dr$ such that $|R|=r$ and $\dim([\mathbf{0}]_{R})=i$, in other words, $a^i_r=0$ for $r\geq n-k+i+1$. 
\end{IEEEproof}

\begin{cor}\label{c1l2t3}
$a^{i}_{n-1}=0$ for every $i\neq k-1$ and  $a^{k-1}_{n-1}=n$.
\end{cor}

\begin{IEEEproof}
Considering $r=n-1 = n-k+(k-1)$ and $i < k-1$, Proposition \ref{P35} ensures  $a^{i}_{n-1}=0$. From Proposition \ref{P33} it follows that  $a^{k}_{n-1}= a^{k}_{d_{k}-1}=0$ and as a particular case of expression  (\ref{sumsum}) it follows that 
\begin{eqnarray} \label{aaa}
\sum_{i=0}^{k}{a^{i}_{n-1}}=\binom{n}{n-1}\text{,}
\end{eqnarray}
and, since $a^{k-1}_{n-1}$ is the only non-zero summand in the equality (\ref{aaa}), it implies $a^{k-1}_{n-1}=\binom{n}{n-1}=n$.
\end{IEEEproof}

\section{Bounds for $P_{\ast}$}

In this section, we establish bounds for $P_{\ast}$ by founding bounds for the coefficients $Q_{\ast,r}$ defined in equality (\ref{defqast}). We start with three lemmas that give us values and bounds for $|[\mathbf{0}]_{R}|$.

\begin{lema}\label{l5t2}
Let  $C$ be an $[n,k]_q$-linear code and let $D\subseteq C$ be an $i$-dimensional linear subcode of $C$. If $\text{supp}(D)\subseteq R \subseteq \dl n \dr$, then $|[\mathbf{0}]_{R}|\geq q^i$. 
\end{lema}

\begin{IEEEproof}
Since $D\subseteq [\mathbf{0}]_{R}$, it follows that  $|[\mathbf{0}]_{R}|\geq |D| = q^i.$
\end{IEEEproof}

\begin{lema}\label{l6t2} 
Let $C$ be an $[n,k]_{q}$-linear code. If $R =supp(D)$ and $D\in {\cal{A}}^{i}_{d_i}$, then $|[\mathbf{0}]_{R}|=q^i$.
\end{lema}

\begin{IEEEproof}
From item (iii) in Proposition \ref{proposicao1} it is known that $[\mathbf{0}]_{R}$ is a vector subspace of ${\mathbb{F}}^{n}_{q}$ and hence $|[\mathbf{0}]_{R}|$ is a power of $q$. We assume that $|[\mathbf{0}]_{R}|\geq q^{i+1}$ and this will lead us to a contradiction. Indeed, $|[\mathbf{0}]_{R}|\geq q^{i+1}$ implies $\dim([\mathbf{0}]_{R})\geq i+1$ hence there is a subspace $D\subseteq [\mathbf{0}]_{R}$ with $\dim(D)=i+1$. From  $D\subseteq [\mathbf{0}]_{R}$ it follows that
\begin{equation*}
supp(D)\subseteq supp([\mathbf{0}]_{R}) \subseteq R
\end{equation*}
hence
\begin{equation*}
d_{i+1} \leq |supp(D)|\leq |supp([\mathbf{0}]_{R})| \leq |R|=d_{i}.
\end{equation*}
But this contradicts the fact  $d_{i}<d_{i+1}$, ensured by the Monotonicity Theorem (Section \ref{teoMONO}). It follows that $|[\mathbf{0}]_{R}|\leq q^{i}$ and Lemma  {\ref{l5t2}} ensures $|[\mathbf{0}]_{R}|=q^i.$
\end{IEEEproof}

In the previous lemma we considered a subset $R\subseteq \llbracket n\rrbracket$ that is the support of a subcode $D\in {\cal{A}}^{i}_{d_i}$ realizing the $i$-th weight. In the following proposition we assume that $|R|=d_i$ but $R\neq supp(D)$ for any $D$ realizing the $i$-th weight.

\begin{lema}\label{l7t2} 
Let $C$ be an  $[n,k]_{q}$-linear code. If $R\subseteq \llbracket n\rrbracket$ satisfies $|R|=d_{i}$ but $R \neq supp(D)$ for any $D\in {\cal{A}}^{i}_{d_i}$, then $|[\mathbf{0}]_{R}|\leq q^{i-1}$.
\end{lema}

\begin{IEEEproof}
Suppose  $|[\mathbf{0}]_{R}|\geq q^i$, or equivalently, $\dim([\mathbf{0}]_{R})\geq i$. In this case, there is an $i$-dimensional $D\subseteq [\mathbf{0}]_{R}$ of $C$ such that $supp(D)\subseteq R$ and 
\begin{equation*}
d_{i}\leq |supp(D)|\leq |supp([\mathbf{0}]_{R})|\leq |R|=d_{i},
\end{equation*}
where the first inequality follows from the minimality of $d_i$, the second one from the fact that $D\subseteq [\mathbf{0}]_{R}$ and the last one from item (iii) in Proposition \ref{proposicao1}. These inequalities imply that	$supp(D)=R$,  $\dim(D)=i$ and $|supp(D)|=d_{i}$, contradicting the hypothesis that $R \neq supp(D)$ for any $D\in {\cal{A}}^{i}_{d_i}$. So,  $|[\mathbf{0}]_{R}| < q^i$ and hence  $|[\mathbf{0}]_{R}|\leq q^{i-1}$.
\end{IEEEproof}

Now we are able to establish bounds for $P_{\ast}(C)$. This will be done in the next theorem, that actually establish upper and lower bounds for some of the coefficients $Q_{\ast ,j}$ in expression (\ref{expancao_Q}).


\begin{teorema}[Bounds for $P_{\ast}(C)$]\label{teoBOUNDS}
Let $C$ be an $[n,k]_q$-linear code. Then, 
\begin{itemize}
\item[(a)] For every $i\in \dl k \dr$, 
$$\Scale[0.9]{Q_{dec,d_i} \geq A^{i}_{d_i}\left(1-\frac{1}{q^{i}}\right)+\left(\binom{n}{d_i} - A^{i}_{d_{i}}\right)\left(1-\frac{1}{\max\{1,q^{k-n+d_i}\}}\right);}$$
\item[(b)] For every $i\in \dl k \dr$, 
$$Q_{dec,d_i} \leq A^{i}_{d_i}\left(\frac{q-1}{q^{i}}\right)+\binom{n}{d_i}\left(1-\frac{1}{q^{i-1}}\right);$$
\item[(c)] For every $i\in \dl 2,k \dr$, 
$$\Scale[0.96]{Q_{amb,d_i} \geq A^{i}_{d_i} + \left(\binom{n}{d_i} - A^{i}_{d_i}\right)\left\lceil 1-\frac{1}{\max\{1,q^{k-n+d_i}\}}\right\rceil.}$$
\item[(d)] $Q_{amb,d_1}=A^{1}_{d_1}$.
\end{itemize}
\end{teorema}

\begin{IEEEproof}
\begin{itemize}
\item[(a)] To simplify the notation we write:
$$\Phi_{i} = \{R \subseteq \dl n \dr; R = supp(D)\ \text{with}\ D\in {\cal{A}}^{i}_{d_i}\}$$
and
$$\tilde{\Phi}_{i} = \{R \subseteq \dl n \dr; |R| = d_{i}\} \setminus \Phi_i.$$
Using this notation and expression  (\ref{defQdec}), the coefficient $Q_{dec,d_i}$ is expressed as 
\begin{equation}\label{eq1teoBOUNDS} 
\Scale[0.93]{Q_{dec,d_i} =\sum_{R\in \Phi_i}{\left( 1-\frac{1}{|[\mathbf{0}]_{R}|}\right)}+\sum_{R\in \tilde{\Phi}_i}{\left( 1-\frac{1}{|[\mathbf{0}]_{R}|}\right)}}.
\end{equation}
Lemma \ref{l6t2} ensures that $R\in \Phi_i$ implies  $|[\mathbf{0}]_{R}|=q^i$ so 
\begin{equation*} 
\Scale[0.95]{Q_{dec,d_i} \geq \sum_{R\in \Phi_i}{\left( 1-\frac{1}{q^i}\right)}+\sum_{R\in \tilde{\Phi}_i}{\left( 1-\frac{1}{|[\mathbf{0}]_{R}|}\right)}}\text{.}
\end{equation*}
Corollary \ref{c8t2} ensures that if $R\in \tilde{\Phi}_i$ then $|[\mathbf{0}]_{R}|\geq q^{k-n+d_i}$ and since  $1-\frac{1}{|[\mathbf{0}]_{R}|}\geq 0$ (for it represents a probability), it follows that
\begin{equation*} 
\Scale[0.93]{Q_{dec,d_i} \geq \sum_{R\in \Phi_i}{\left( 1-\frac{1}{q^i}\right)}+\sum_{R\in \tilde{\Phi}_i}{\left( 1-\frac{1}{\max\{1,q^{k-n+d_i}\}}\right)}}.
\end{equation*}
Lemma \ref{L32} implies that $|\Phi_i|=A^{i}_{d_i}$ and since the summands do  not depend on $R$ we get that 

\begin{equation*}
\Scale[0.9]{Q_{dec,d_i}\geq A^{i}_{d_i}\left(1-\frac{1}{q^{i}}\right)+\left(\binom{n}{d_i} - A^{i}_{d_{i}}\right)\left(1-\frac{1}{\max\{1,q^{k-n+d_i}\}}\right).}
\end{equation*}

\item[(b)] From Lemma  \ref{l7t2} it is possible to bound $|[\mathbf{0}]_R|\leq q^{i-1}$, for $R \in \tilde{\Phi}_i$,  and  Lemma \ref{l6t2} establishes an expression to  $|[\mathbf{0}]_{R}|$, for $R \in \Phi_i$. Substituting those values in expression (\ref{eq1teoBOUNDS}) it follows that 
\begin{equation*} 
Q_{dec,d_i} \leq A^{i}_{d_i}\left(\frac{q-1}{q^{i}}\right)+\binom{n}{d_i}\left(1-\frac{1}{q^{i-1}}\right).
\end{equation*}
 
\item[(c)] From Lemma \ref{l6t2} it follows that
\begin{equation}\label{eq2tt}
\left\lceil  1-\frac{1}{|[\mathbf{0}]_{R}|}\right\rceil = 1
\end{equation}
and, from Corollary \ref{c8t2}, 
\begin{equation}\label{eq3tt}
\left\lceil  1-\frac{1}{|[\mathbf{0}]_{R}|}\right\rceil \geq  \left\lceil 1-\frac{1}{\max\{1,q^{k-n+d_i}\}}\right\rceil .
\end{equation}
Expression (\ref{defQamb}) may be written as
\begin{equation}
\Scale[0.9]{Q_{dec,d_i} =\sum_{R\in \Phi_i}{\left\lceil  1-\frac{1}{|[\mathbf{0}]_{R}|}\right\rceil}+\sum_{R\in \tilde{\Phi}_i}{\left\lceil  1-\frac{1}{|[\mathbf{0}]_{R}|}\right\rceil}}, 
\end{equation}
and substituting it into  (\ref{eq2tt}) and (\ref{eq3tt}) it follows that 
\begin{equation}
\Scale[0.9]{Q_{amb,d_i}\geq A^{i}_{d_i} + \left(\binom{n}{d_i} - A^{i}_{d_i}\right)\left\lceil 1-\frac{1}{\max\{1,q^{k-n+d_i}\}}\right\rceil.}
\end{equation}

\item[(d)] Expression (\ref{defqast}) implies that 
$$Q_{amb,d_1} = \sum_{i=0}^{k}a^{i}_{d_1} \left\lceil 1-q^{-i}\right\rceil. $$
From Corollary \ref{pequeno}, only two of the summands above are non zero, namely
\begin{align*}
Q_{amb,d_1} & = a^{0}_{d_1}\left\lceil 1-q^{-0}\right\rceil + a^{1}_{d_1}\left\lceil 1-q^{-1}\right\rceil\\
 & = a^{1}_{d_1}\left\lceil 1-q^{-1}\right\rceil
\end{align*}
and from Proposition \ref{P33} it follows that 
$$Q_{amb,d_1} = a^{1}_{d_1} = A_{d_1}^{1}.$$

\end{itemize}
\end{IEEEproof}

\section{$P_{\ast}$ and separability properties}\label{MDS}

We start this section presenting some separability properties that generalize the concept of MDS codes and then we will study the behavior of the bounds for $P_{\ast}(C)$ expressed in  Theorem \ref{teoBOUNDS} for codes having some of those separability properties. 

The Singleton bound states that $d_{1}(C)\leq n-k+1 $ and a code that satisfies this bound is said to be \emph{Maximum Distance Separable} (MDS). The \emph{ Singleton defect} $s(C)$ of an $[n,k]_{q}$-code  $C$ is the measure of how much apart from being MDS a code is:  
$$s(C)=n-k+1-d_{1}(C).$$
Using the defect, we say that a code $C$ is \emph{Maximum Distance Separable} (MDS) if $s(C)=0$ and (following Boer in \cite{boe}) $C$ is \emph{Almost Maximum Distance Separable} (AMDS) if $s(C)=1$.

Considering the generalized weights, there are more then one reasonable way to generalize and express the separability property of a linear code. Considering the monotonicity of the weight hierarchy,  the $i$\emph{-th Singleton defect} of an  $[n,k]_{q}$-linear code $C$ is defined as
$$s_{i}(C)=n-k+i-d_{i}(C).$$
Following Wei (in \cite{wei}) a code $C$ is said to be a  \emph{$j$-MDS code} if $s_{j}(C)=0$. If $s_{j}(C)=1$, the code is said to be a \emph{$j$-AMDS code}.

We say that $C$ is a \emph{proper $j$-MDS} code (or just \emph{$P_{j}$-MDS}) if it is $j$-MDS and proper in the sense that 
$$j=\min\{i\in\llbracket k\rrbracket; C\ \text{ is an }\ i\text{-MDS code} \}.$$ 

Similarly, we say that $C$ is an \emph{ $P_{j}$-AMDS code} if
$$j=\min\{i\in\llbracket k\rrbracket; C\ \text{ is an }\ i\text{-AMDS code}\}.$$

\subsection{Expressions for $P_{amb}(C)$ and $P_{dec}(C)$}
We consider the matrix $\Lambda$ used in  Proposition \ref{eq8}  to give a vectorial expression for the ambiguity or the decoding error probability $P_{\ast}$.  Propositions  $\ref{P33}$ and $\ref{P35}$ ensure that many of the coefficients of $\Lambda^{T}$ are null. Let us write $\Lambda^{T}$ explicitly as:

\begin{frame}
\footnotesize
\arraycolsep=3pt 
\medmuskip = 1mu 

\begin{displaymath}
\Lambda^{T}=\left( \begin{array}{ccccc}
\textcolor{blue}{1}     & \textcolor{blue}{0}      & \textcolor{blue}{\cdots} & \textcolor{blue}{0}       & \textcolor{blue}{0} \\
\textcolor{red}{\vdots} & \textcolor{blue}{\vdots} &                          &  \textcolor{blue}{\vdots} & \textcolor{blue}{\vdots} \\
\textcolor{red}{\binom{n}{d_{1}-1}} & \textcolor{blue}{0} & \textcolor{blue}{\cdots} & \textcolor{blue}{0} & \textcolor{blue}{0} \\
\textcolor{red}{\binom{n}{d_{1}}- A^{1}_{d_{1}}} & \textcolor{blue}{A^{1}_{d_{1}}} & \textcolor{blue}{\cdots} &\textcolor{blue}{0} & \textcolor{blue}{0} \\
a^{0}_{d_{1}+1} & a^{1}_{d_{1}+1} & \textcolor{blue}{\cdots} & \textcolor{blue}{0} & \textcolor{blue}{0} \\


\vdots & \vdots & & \textcolor{blue}{\vdots}  & \textcolor{blue}{\vdots} \\
a^{0}_{d_{k-1}-1} & a^{1}_{d_{k-1}-1} & \cdots & \textcolor{blue}{0} &\textcolor{blue}{0} \\
a^{0}_{d_{k-1}} & a^{1}_{d_{k-1}} & \cdots &\textcolor{blue}{A^{k-1}_{d_{k-1}}} &\textcolor{blue}{0} \\
a^{0}_{d_{k-1}+1} & a^{1}_{d_{k-1}+1} & \cdots & a^{k-1}_{d_{k-1}+1}   &\textcolor{blue}{0}   \\
\vdots         & \vdots               &      & \vdots  & \textcolor{blue}{\vdots} \\
a^{0}_{n-k} & a^{1}_{n-k} &   \cdots &  a^{k-1}_{n-k} & \textcolor{blue}{0}\\
\textcolor{green}{0} & a^{1}_{n-k+1} &   \cdots &  a^{k-1}_{n-k+1} & \textcolor{blue}{0}\\
\textcolor{green}{0} & \textcolor{green}{0} &   & a^{k-1}_{n-k+2}  & \textcolor{blue}{0}\\
\textcolor{green}{0} & \textcolor{green}{0} &  & \textcolor{blue}{\vdots} & \textcolor{blue}{\vdots}\\
\textcolor{green}{\vdots} & \textcolor{green}{\vdots} & \textcolor{green}{\ddots} & a^{k-1}_{n-2} & \textcolor{blue}{0}\\
\textcolor{green}{0} & \textcolor{green}{0}  & \textcolor{green}{\cdots} & \textcolor{red}{n} & \textcolor{blue}{0}\\

\textcolor{green}{0} & \textcolor{green}{0} &  \textcolor{green}{\cdots} & \textcolor{green}{0}& \textcolor{blue}{A^{k}_{d_{k}}}
\end{array} \right)
\end{displaymath}

\end{frame}

In this presentation, the values of the blue entries are established in Proposition  \ref{P33} and the green entries by Proposition $\ref{P35}$. Looking at expression (\ref{sumsum}), we see it  sums over the lines of $\Lambda^{T}$. There are three lines for which only one entry is unknown (neither blue nor green) and these entries may be determined  from the sum (\ref{sumsum}): those are the three entries in red.

Looking now at the columns of $\Lambda^{T}$, we see that the quantity of undetermined entries at the column $j$ is given by the difference $(n-k+i)-d_i$ and, in an informal way, we can state that ``\textit{the more }$C$ \textit{is separable, the more the entries of }$\Lambda^{T}$ \textit{are known}". In particular, assuming that $C$ is MDS, that is, that $d_1=n-k+1$, the monotonicity of the weights implies that $d_i = n-k+i$ for every $i\in \dl k \dr$ and in this case, all nonzero entries of $\Lambda$ are expressed in terms of the weight spectra, namely, in terms of $A^1_{d_1},A^2_{d_2},\cdots ,A^k_{d_k}$. But for an MDS code, the following theorem (due to Han, in \cite{dou1}) gives explicit expressions for those coefficients, depending exclusively on $n$, $k$ and $q$:

\begin{teorema}[Theorem 2.5 in \cite{dou1}]\label{teodou}
Let $C$ be an $[n,k]_{q}$-linear code and suppose that $C$ is $P_{s}$-MDS. Then, for $s \leq i \leq k$, we have that 
$$A^{i}_{r}(C)=\left\{\begin{array}{cc}
0, & \text{if}\ 0\leq r \leq d_{i} \\
\binom{n}{r}\sum^{r-d_{i}}_{t=0}(-1)^{t}\binom{r}{t}\genfrac[]{0pt}{}{r+i-d_{i}-t}{i}_q, & \text{if}\ d_{i}< r \leq n
\end{array}\right.$$
\end{teorema}

Here, $\genfrac[]{0pt}{}{j}{i}_q$ is the Gaussian binomial coefficient and $P_{s}$-MDS stands for \textit{proper} $s$-MDS, that is, the code $C$ has Singleton defect equals to $s$. Using those expressions for $A^1_{d_1},A^2_{d_2},\cdots ,A^k_{d_k}$ we have an alternative proof of the following Theorem (already proved by Kasami and Lin in \cite{kasami}):

\begin{teorema}\label{e1l1t1} 
Let $C$ be an MDS code. Then,  
\begin{itemize}
\item[(a)] $P_{amb}(C) =\sum_{i=n-k+1}^{n}\binom{n}{i}\mathtt{p}^{i}(1-\mathtt{p})^{n-i}$;
\item[(b)] $P_{dec}(C) =\sum_{i=n-k+1}^{n}\binom{n}{i}\left( 1-\frac{1}{q^{i}}\right)\mathtt{p}^{i}(1-\mathtt{p})^{i}$.
\end{itemize}
\end{teorema}

If $C$ is an AMDS code, that is, if  $d_1=n-k$, there is an unique $s:=s(C)\leq k$ such that $C$ is $P_{s}$-MDS and we can determine an explicit formula for $P_{\ast}(C)$ depending only on $A_{d_1}^{1}$ and $s$:

\begin{teorema}\label{t6} 
Let C be an $[n,k]_{q}$ AMDS linear code and let $s:=s(C)\leq k$ such that $C$ is $P_{s}$-MDS. Then,
\begin{align*}
P_{amb}(C) =&A^{1}_{n-k}{\mathtt{p}}^{n-k}(1-\mathtt{p})^{k}\\
+&\sum_{i=1}^{k}\binom{n}{n-k+i}{\mathtt{p}}^{n-k+i}(1-\mathtt{p})^{k-i}
\end{align*}
and
\begin{align*}
P_{dec}(C) =& \sum_{i=0}^{s-2}A^{i+1}_{n-k+i}\left(\frac{q-1}{q^{i+1}}\right){\mathtt{p}}^{n-k+i}(1-\mathtt{p})^{k-i}\\ 
+&\sum_{i=0}^{k}\binom{n}{n-k+i}\left(1-\frac{1}{q^{i}}\right){\mathtt{p}}^{n-k+i}(1-\mathtt{p})^{k-i}.
\end{align*}

\end{teorema}

\begin{IEEEproof}
It follows straightforward from the use of the vectorial form (\ref{eq8}) and Propositions \ref{P33} and \ref{P35}.
\end{IEEEproof}


We remark that in the proof of Theorem \ref{t6}, we actually considered that in the sum (\ref{defqast})  that expresses $Q_{\ast,r}$, the coeficients $a_r^ i$ are zero for many values, as we can see in Proposition \ref{P35}. Indeed, if a code has Singleton defect $s$ (hence it is $P_s$-MDS), then all the $A_r^i$ are zero for $r\neq d_i$ and for $i\geq s$. This same approach can be used to improve the bounds given by Liva et. al in \cite{liva}. 

We start with the following:

\begin{prop}
\label{t5} Let $C$ be a ${P}_{s}$-MDS $[n,k]_{q}$-code . Then
\[
Q_{dec,r}=\binom{n}{r}\left(  1-\frac{1}{q^{r-n+k}}\right)  \text{ and }  Q_{amb,r}=\binom{n}{r}
\]
for $d_{s}\leq r\leq n$.
\end{prop}

\begin{IEEEproof}
First of all we recall that we are assuming every code to be irreducible, in the sense that $d_k=n$. 
Since $d_{k}=n$ and $d_{s}=n-k+s$ (for $C$ is assumed to be $P_s$-MDS), it follows that 
$\llbracket d_{s},d_{k}\rrbracket$ has  $k-s+1$ elements and the Monotonicity Theorem (Section \ref{teoMONO}) ensures that 
\[
\llbracket d_{s},d_{k}\rrbracket=\{d_{i};i\in\llbracket s,k\rrbracket\}.
\]
So, if  $d_{s}\leq r\leq n$ then $r=d_{j}=n-k+j$ for some $j\in\llbracket s,k\rrbracket$ and Theorem 
\ref{teodou} ensures that
\begin{equation}\label{e1}
A_{d_{j}}^{j}=\binom{n}{d_{j}}
\end{equation}
for every $j\in\llbracket s,k\rrbracket$. The first two items in Theorem  \ref{teoBOUNDS} 
ensures that
\begin{equation}\label{e2}
\Scale[0.95]{A_{d_{j}}^{j}\left(  1-\frac{1}{q^{j}}\right)  \leq Q_{dec,d_{j}}\leq
A_{d_{j}}^{j}\left(  \frac{q-1}{q^{j}}\right)  +\binom{n}{d_{j}}\left(
1-\frac{1}{q^{j-1}}\right)}
\end{equation}
for every $j\in\llbracket s,k\rrbracket$. Substituting (\ref{e1}) into
(\ref{e2}) we find that the lower and the upper bounds coincide and hence 
\[
Q_{dec,r}=\binom{n}{r}\left(  1-\frac{1}{q^{r-n+k}}\right)
\]
for every $r\in\llbracket d_{s},d_{k}\rrbracket$. 

Looking now at $Q_{amb,d_{j}}$, from Theorem \ref{teoBOUNDS} it follows that 
\begin{equation}\label{e3}
A_{d_{j}}^{j}\leq Q_{amb,d_{j}}\leq\binom{n}{d_{j}},
\end{equation}
for every $j\in\llbracket2,k\rrbracket$. So, if $s\geq2$, substituting
(\ref{e1}) in (\ref{e3}) we find again that the lower and the upper bounds coincide and hence 
\[
Q_{amb,r}=\binom{n}{r},
\]
for every $r\in\llbracket d_{s},d_{k}\rrbracket$. For the remaining case, $s=1$, from Theorem
\ref{teoBOUNDS} and identity (\ref{e1}) it follows that
\[
Q_{amb,d_{1}}=|\Phi_{1}|=A_{d_{1}}^{1}=\binom{n}{d_{1}}\text{.}
\]
\end{IEEEproof}

As an example of how it is possible to improve the results in \cite{liva}, 
Liva et al.  give an  upper bound for the error probability (Theorem 4 in \cite{liva}) as an expression that involves a sum of $n-k$ terms that are added to the Singleton bound $P_{B}^{S}\left(  n,k,\mathtt{p}\right)$:

\begin{align*}
P_{dec}\left(  C\right)    & \leq P_{B}^{S}\left(  n,k,\mathtt{p}\right)  \\
& \Scale[0.85]{ +\sum
_{i=1}^{n-k}\dbinom{n}{i}\mathtt{p}^{i}\left(  1-\mathtt{p}\right)  ^{n-i}\min\left\{
1,\frac{1}{q-1}\sum_{j=1}^{i}\dbinom{i}{j}\frac{A_{j}^{1}}{\dbinom{n}{j}%
}\right\} } \\
& =P_{B}^{S}\left(  n,k,\mathtt{p}\right)  \\
& \Scale[0.85]{+\sum_{\mathbf{i=d}_{1}}^{n-k}\dbinom{n}%
{i}\mathtt{p}^{i}\left(  1-\mathtt{p}\right)  ^{n-i}\min\left\{  1,\frac{1}{q-1}\sum_{j=1}%
^{i}\dbinom{i}{j}\frac{A_{j}^{1}}{\dbinom{n}{j}}\right\}}\text{.}
\end{align*}
If $C$ is $P_{s}$-MDS we have that $Q_{\ast,i}=\dbinom{n}{r}\left(
1-\frac{1}{q^{r-n+k}}\right)  $ for $i\geq d_{s}$ and this is the coefficient of
the Singleton bound $P_{B}^{S}\left(  n,k,\mathtt{p}\right)  $. Hence 
those terms may be omitted from the sum and  the bound can improved:%
\begin{align*}
P_{dec}\left(  C\right)    & \leq P_{B}^{S}\left(  n,k,\mathtt{p}\right) \\
&\Scale[0.85]{+\sum_{i=d_{1}}^{\mathbf{d_{s}}}\dbinom{n}{i}\mathtt{p}^{i}\left(  1-\mathtt{p}\right)  ^{n-i}%
\min\left\{  1,\frac{1}{q-1}\sum_{j=1}^{i}\dbinom{i}{j}\frac{A_{j}^{1}%
}{\dbinom{n}{j}}\right\} } \\
& \leq P_{B}^{S}\left(  n,k,\mathtt{p}\right)  \\
&\Scale[0.85]{+\sum_{i=d_{1}}^{n-k}\dbinom{n}{i}%
\mathtt{p}^{i}\left(  1-\mathtt{p}\right)  ^{n-i}\min\left\{  1,\frac{1}{q-1}\sum_{j=1}%
^{i}\dbinom{i}{j}\frac{A_{j}^{1}}{\dbinom{n}{j}}\right\}}
\end{align*}
where the second inequality is strict if $d_{s}<n-k$.

Theorem \ref{t6} ensures that, for an AMDS code, the ambiguity probability $P_{amb}(C)$ is determined by $A^1_{n-k}$ and the decoding error probability $P_{dec}(C)$ is completely determined by  the coefficients $A^1_{n-k}$, $A^1_{n-k}, \cdots , A^{s-1}_{n-k+s-2} $ of the spectra-matrix. It follows that bounds for the coefficients of the spectra-matrix leads to bounds for $P_{\ast}(C)$. In the particular case of an NMDS-code (\textsl{near MDS}, a code $C$ such that $d_1(C)=n-k$ and $d_2(C)=n-k+2$), the coefficient $A^1_{n-k}$ fully determines $P_{\ast}(C)$ and an upper bound for this coefficient is provided by Dodunekov and Landgev, in \cite{dodu}: $A^1_{n-k} \leq \binom{n}{k-1}\frac{q-1}{k}$.

\begin{figure}[h]
	\centering
		\includegraphics[scale=0.62]{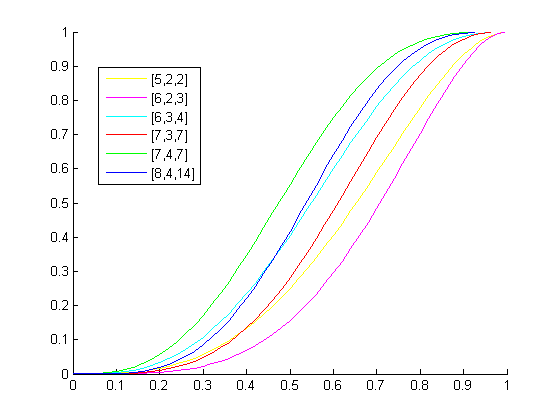}
	\caption{Ambiguity probability of AMDS codes}
	\label{fig:garfico}
\end{figure}

If we consider, for example,  binary systematic AMDS code of minimum distance at least three and cardinality at least four, those codes were recently classified in \cite{RAVAGNANI}[Theorem 19]: There are exactly six such codes, with parameters $\left[n,k,A_{n-k}^1\right]$ given by $\left[5,2,2\right]$,  $\left[6,2,3\right]$, $\left[6,3,4\right]$, $\left[7,3,7\right]$, $\left[7,4,7\right]$, and $\left[8,4,14\right]$.
The ambiguity probability $P_{amb}$ of each of those codes (as a function of the overall error probability $\mathtt{p}$) is pictured in Figure \ref{fig:garfico}.

We remark that codes may have different behavior for different values of $\mathtt{p}$, so that we do have crossing lines in Figure \ref{fig:garfico}. In the next section we take into account the overall error probability  $\mathtt{p}$ and show the minimizing property of MDS and AMDS codes for small values of $\mathtt{p}$.

\subsection{Behavior of $P_{\ast}$ for small $\mathtt{p}$ and optimality of MDS and AMDS codes}\label{ppequeno}

As expected, for small overall error probability $\mathtt{p}$, minimizing error probability $P_{*}(C)$ demands to maximize $d_1(C)$: 
\begin{prop}\label{p1t7}
Let $C_{1}$ and $C_{2}$ be two $[n,k]_{q}$-linear codes. For $\mathtt{p}$ sufficiently small, if $d_{1}(C_{1})>d_{1}(C_{2})$ then $P_{*}(C_{1})<P_{*}(C_{2})$.
\end{prop}

\begin{IEEEproof}

To prove the proposition we assume $d_{1}(C_{1})>d_{1}(C_{2})$ and show that 
$$\lim_{\mathtt{p}\rightarrow 0}{\frac{P_{*}(C_{1})}{P_{*}(C_{2})}}=0 .$$
Considering the expansion $P_{\ast}(C) =\sum_{r=0}^{n} Q_{\ast,r} \mathtt{p}^{r} (1-\mathtt{p})^{n-r}$  obtained in equation (\ref{expancao_Q}), we have that
\begin{align*}\lim_{\mathtt{p}\rightarrow 0}{\frac{P_{*}(C_{1})}{P_{*}(C_{2})}} &= \lim_{\mathtt{p}\rightarrow 0}\frac{\sum_{i=d_{1}(C_{1})}^{n}Q_{*,i}(C_{1}){\mathtt{p}}^{i}(1-\mathtt{p})^{n-i}}{\sum_{j=d_{1}(C_{2})}^{n}Q_{*,j}(C_{2}){\mathtt{p}}^{j}(1-\mathtt{p})^{n-j}}\\ &=\lim_{\mathtt{p}\rightarrow 0}\frac{\sum_{i=d_{1}(C_{1})}^{n}Q_{*,i}(C_{1})\left(\frac{\mathtt{p}}{(1-\mathtt{p})}\right)^{i}}{\sum_{j=d_{1}(C_{2})}^{n}Q_{*,j}(C_{2})\left(\frac{\mathtt{p}}{(1-\mathtt{p})}\right)^{j}}\text{.}
\end{align*}

Denoting $x = \frac{\mathtt{p}}{(1-\mathtt{p})}$ and noting that $\lim_{p\rightarrow 0}\frac{\mathtt{p}}{(1-\mathtt{p})}=0$ it follows that

\begin{align*}\label{limix0}
\lim_{x\rightarrow 0}{\frac{P_{*}(C_{1})}{P_{*}(C_{2})}} &= \lim_{x\rightarrow 0}\frac{\sum_{i=d_{1}(C_{1})}^{n}Q_{*,i}(C_{1})x^{i}}{\sum_{j=d_{1}(C_{2})}^{n}Q_{*,j}(C_{2})x^{j}}\\
&=\lim_{x\rightarrow 0}\frac{\sum_{i=d_1(C_1)-d_1(C_2)}^{n-d_1(C_2)}Q_{*,i+d_1(C_2)}(C_{1})x^{i}}{\sum_{j=0}^{n-d_1(C_2)}Q_{*,j+d_1(C_2)}(C_{2})x^{j}}
\end{align*}

and since   $d_{1}(C_{1})-d_{1}(C_{2})>0$, we have that
$$\lim_{x\rightarrow 0}{\frac{P_{*}(C_{1})}{P_{*}(C_{2})}}= 0 < 1\text{.}$$
Considering the quotient $\frac{P_{*}(C_{1})}{P_{*}(C_{2})}$ as a function of overall error probability, it depends continuously on $\mathtt{p}$ and so, since its limit is $0$ it follows that $\frac{P_{*}(C_{1})}{P_{*}(C_{2})}<1$  for $\mathtt{p}<\mathtt{p_0}$ for some $\mathtt{p_0}$, or equivalently, $P_{*}(C_{1})<P_{*}(C_{2})$, for every $\mathtt{p}$ sufficiently small.
\end{IEEEproof}

If for a given pair $(n,k)$ there exist an  MDS (AMDS) $[n,k]_q$-code, we say that the triple $(n,k,q)$ is an  \emph{MDS} (\emph{AMDS}) triple. As an immediate consequence of Proposition \ref{p1t7} we have the following proposition (already known and proved in \cite{Fas}):
\begin{prop} If $(n,k,q)$ is MDS and $C$ is an $[n,k]_q$-code that minimizes the error probability, then $C$ is MDS.
\end{prop}

Triples that are MDS are not very frequent . For $q=2$, for example, it is well known that MDS-codes are rather trivial and the unique MDS triples are  $(n,n,2)$, $(n,1,2)$, $(n,0,2)$, $(n,n-1,2)$. Considering AMDS codes, those are not  classified, but there are many constructions of particular families of AMDS codes and results ensuring the existences of such codes with parameters $n$ and $k$ (see for example  \cite{boe}). In all those cases, when the triple $(n,k,q)$ is AMDS but not MDS, for $\mathtt{p}$ sufficiently small, a code that minimizes $P_{\ast }$ must be an AMDS code.

Proposition \ref{p1t7} states that for $\mathtt{p}$ sufficiently small, we should look for codes having maximal minimal distance. Among all those codes with the same (maximal) minimal distance, which should perform better? A partial answer is given by the next two results and can be summarized as follows: maximize the minimal distance and then minimize the corresponding value in the spectra.

\begin{prop}\label{p2t7}
Let $C_{1}$ and $C_{2}$  be two $[n,k]_{q}$-linear codes with $d_{1}(C_{1})= d_{1}(C_{2})$. If $Q_{*,d_{1}(C_{1})}<Q_{*,d_{1}(C_{2})}$, then $P_{*}(C_{1})<P_{*}(C_{2})$, for $\mathtt{p}$ sufficiently small.
\end{prop}

\begin{IEEEproof}
From equation (\ref{expancao_Q}) it follows that
$$\lim_{x\rightarrow 0}{\frac{P_{*}(C_{1})}{P_{*}(C_{2})}}= \lim_{x\rightarrow 0}\frac{\sum_{i=d_{1}(C_{1})}^{n}Q_{*,i}(C_{1})x^{i}}{\sum_{i=d_{1}(C_{2})}^{n}Q_{*,i}(C_{2})x^{i}}.$$
We write $d_1=d_1(C_1)=d_1(C_2)$ and cancel  $x^d_1$ from the right side we get
\begin{align*}
\lim_{x\rightarrow 0}{\frac{P_{*}(C_{1})}{P_{*}(C_{2})}} &= \lim_{x\rightarrow 0}\frac{Q_{*,d_{1}}(C_1)+\sum_{i=d_{1}+1}^{n}Q_{*,i}(C_{1})x^{i-d_{1}}}{Q_{*,d_{1}}(C_2)+\sum_{i=d_{1}+1}^{n}Q_{*,i}(C_{2})x^{i-d_{1}}}\\
&= \frac{Q_{*,d_{1}}(C_1)}{Q_{*,d_{1}}(C_2)}<1,
\end{align*}
hence $P_{*}(C_{1})<P_{*}(C_{2})$ for every $\mathtt{p}$ sufficiently small.
\end{IEEEproof}

\begin{prop}\label{C20}
Let $C_{1}$ and $C_{2}$  be two $[n,k]_{q}$-linear codes with $d_{1}(C_{1})= d_{1}(C_{2})$. If $A^{1}_{d_{1}}(C_{1})<A^{1}_{d_{1}}(C_{2})$, then $P_{*}(C_{1})<P_{*}(C_{2})$, for $\mathtt{p}$ sufficiently small. 
\end{prop}

\begin{IEEEproof}
It follows straightforward from  Proposition \ref{p2t7} and items \textit{(b)} and \textit{(d)} of Theorem \ref{teoBOUNDS}.
\end{IEEEproof}

\section{Conclusion}
In this work we used the generalized weights and spectra to set new bounds for the error probability over an erasure channel. Further work may be done exploring the situation when two codes have the same minimal distance and this is attained by the same number of vectors. The role of generalized weights and spectra  for the error probability still needs to be explained for other channels.


\section*{Acknowledgment}
The first author was partially supported by CAPES and the second was partially supported by grants 2013/25977-7 and 2013/09493-0, S\~{a}o Paulo Research Foundation (FAPESP). 



%

\bibliographystyle{IEEEtran}
\bibliography{TESE25042013}

\end{document}